\documentclass[sigconf]{acmart}

\usepackage{booktabs}
\usepackage{colortbl}
\usepackage{graphicx}
\usepackage{tikz}
\usepackage{multirow}
\usetikzlibrary{arrows.meta,positioning}
\colorlet{sigA}{black!10}  
\colorlet{sigB}{black!6}   
\colorlet{sigC}{black!3}   
\graphicspath{{figures/}}

\AtBeginDocument{%
  }

\setcopyright{none}
\renewcommand\footnotetextcopyrightpermission[1]{}
\begin{document}

\title{Regime-Gated Residual Mixture-of-Experts for Cross-Sectional Volatility
Forecasting}


\author{Junyi Ye}
\orcid{0000-0002-6348-5207}
\affiliation{%
  \institution{School of Computing, Montclair State University}
  \city{Montclair}
  \state{New Jersey}
  \country{USA}
}
\email{yej@montclair.edu}

\author{Gargi Vijay Borde}
\affiliation{%
  \institution{School of Computing, Montclair State University}
  \city{Montclair}
  \state{New Jersey}
  \country{USA}
}
\email{bordeg1@montclair.edu}
\begin{abstract}
Financial volatility is regime dependent, yet incorporating regime
information into neural networks can also destabilize training.
This paper asks where such information should enter a neural
cross-sectional volatility forecasting model. We study five-day realized-volatility forecasts for 1,027 U.S. equities using a rolling walk-forward evaluation framework in which information, model capacity, hyperparameter tuning, and random seeds are matched across architectures.
We propose RG-ResMoE, a regime-gated residual mixture-of-experts
architecture in which regime information is used only for expert
routing rather than for direct forecasting. The base predictor models volatility from stock features, while a gating network uses regime state variables to route residual corrections. RG-ResMoE consistently outperforms a capacity-matched MLP in both forecasting accuracy and training     stability in the main U.S. study. Similar gains are observed on an independent Japanese panel. The integration pathway is decisive: appending the same regime variables directly to the forecasting input degrades both predictive performance and training stability, whereas restricting them to the routing gate improves accuracy and Value-at-Risk calibration. Hard
routing consistently underperforms soft routing.
The results suggest that, in compact neural volatility
forecasting models, the primary value of mixture-of-experts models lies less
in increasing model capacity than in controlling how nonstationary
regime information influences prediction.

\end{abstract}






\pagestyle{plain}

\maketitle

\section{Introduction}

Financial volatility forecasting is central to portfolio allocation,
derivative pricing, and risk management. Deep neural networks have
substantially improved volatility prediction by learning nonlinear
relationships from large cross-sectional equity panels. However,
financial markets are inherently nonstationary: the relationship
between historical observations and future volatility changes across
periods of calm, crisis, and recovery. Regime information is therefore
a natural source of additional context for neural forecasting models.
The central question, however, is not whether market regimes matter,
but how such information should be incorporated into the forecasting
process.

Existing approaches introduce regime information in several ways. The
simplest strategy appends regime variables directly to the model input~\cite{tian2026regime,lee2025graph}.
Regime-switching models instead maintain different predictors for
different market states~\cite{hamilton1989new,hamilton1994autoregressive}, while mixture-of-experts (MoE) architectures
learn observation-dependent routing through a gating network~\cite{he2026raven}. Although
these approaches differ substantially, they typically change the regime
representation, routing mechanism, and model architecture
simultaneously. As a result, it remains unclear \textit{whether performance
differences arise from the regime information itself or from where that
information enters the model}. This distinction is particularly
important in financial forecasting, where predictive gains are often
small, optimization is sensitive to initialization, and additional
model capacity alone does not necessarily improve generalization.

In this paper, we study this question through an information-matched
comparison in which the regime variables, forecasting backbone,
parameter budget, hyperparameter selection, and evaluation protocol
are all held fixed, while only the integration pathway is varied. We
propose \emph{RG-ResMoE}, a residual mixture-of-experts architecture in
which a frozen base network first predicts the volatility level and
small zero-initialized experts learn only residual corrections. A soft
gate observes both stock features and regime variables to determine how
these corrections are combined, while the forecasting networks
themselves receive only stock features. This design separates two
choices that are often confounded: whether regime information is
available, and where it is allowed to influence the prediction.

Experiments on a rolling walk-forward evaluation covering 1,027 U.S.
equities from 2018 to 2025 reveal three consistent findings. First,
the location of regime information is decisive. Appending the same
regime variables directly to the forecasting input consistently
reduces forecasting accuracy and substantially increases training
instability, whereas using them only for gated residual routing
improves both accuracy and robustness. Second, continuous soft routing
consistently outperforms hard routing based on learned assignments or
hand-crafted market partitions, indicating that smooth state-dependent
reweighting is more effective than discrete regime selection. Third,
the gains are concentrated in the market conditions where volatility
forecasts matter most, improving Value-at-Risk calibration and
producing the largest forecasting improvements during periods of
elevated market volatility. The same stability ordering is reproduced
on an independent Japanese equity panel.

More broadly, our results suggest that the primary value of regime-aware MoE models in compact financial forecasting systems lies less in increasing model capacity than in controlling how nonstationary market information influences the prediction. Taken together, these findings establish a simple design principle: in compact neural volatility forecasters, regime information is most effective when it guides residual routing rather than being treated as ordinary predictive input. We believe this principle extends beyond the proposed RG-ResMoE architecture and provides guidance for incorporating market state information into future neural forecasting models.

\begin{figure*}[t]
\centering
\includegraphics[width=0.95\linewidth]{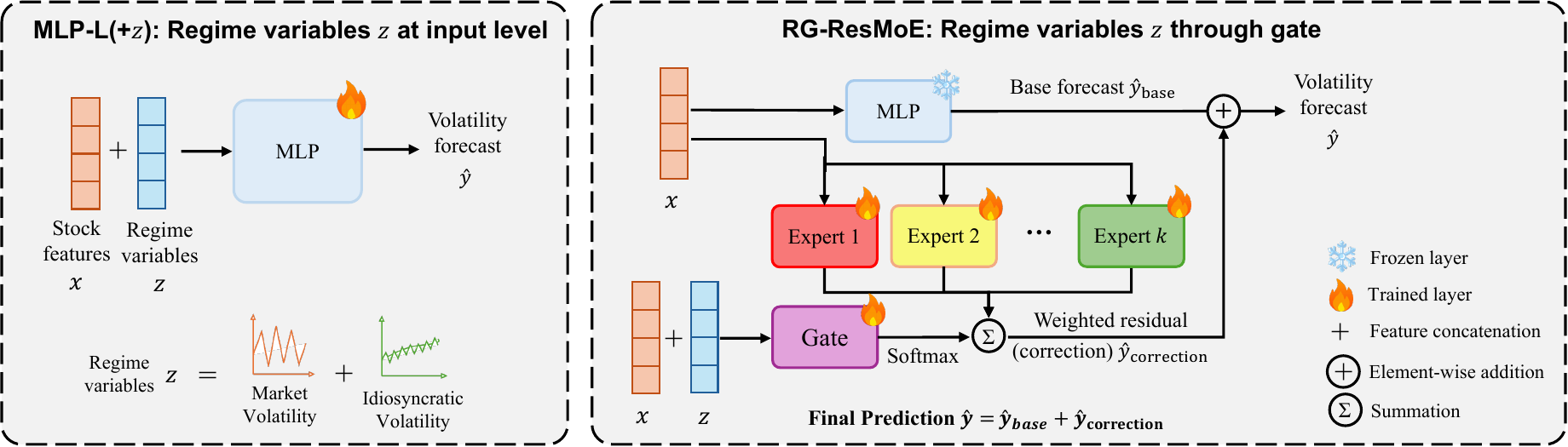}
\caption{Architectures of the baseline MLP-L(+$z$) (left) and proposed RG-ResMoE (right).}
\label{fig:framework}
\end{figure*}

\section{Data and Task}\label{sec:data}

\subsection{Equity panels}

Our main experiments use a U.S. equity panel, with a Japanese
panel reserved for cross-market replication. The U.S. panel contains
1{,}027 common stocks from Yahoo Finance, observed from December
2015 to November 2025. The universe spans all eleven sectors and
large-, mid-, and small-cap stocks. Among these stocks, 85\% are
current S\&P 1500 constituents. The panel is restricted to stocks
that remain listed through the end of the sample and have at least
six years of price history. Among them, 92\% cover the full ten-year
period.

The replication panel contains 1{,}552 Japanese stocks from Yahoo Finance over the same
period and covers the TSE Prime segment. It is constructed using the
same data-processing pipeline and serves as a second-market
replication of the main U.S. results.

\subsection{Inputs and regime state variables}

For each stock $i$ on day $t$, the \textbf{stock-level input}
$x_{i,t}$ contains sixteen features computed from the stock's own
price history. These include realized volatility over the trailing
5, 20, and 60 days, cumulative returns over the trailing 5 and
20 days, the 14-day relative strength index (RSI), and the ten most recent daily log returns.
For each walk-forward window, normalization parameters are estimated
from the training data and then applied to the validation and test data.

The \textbf{regime state variable} $z_{i,t}$ consists of two
volatility-based indicators: market volatility and idiosyncratic
volatility. Market volatility is defined as the 20-day rolling
volatility of the equal-weighted market return $r_{m,t}$ and is shared
by all stocks on a given trading day. Idiosyncratic volatility is
defined as the 20-day rolling volatility of the residual return
$r_{i,t}-\beta_{i,t}r_{m,t}$, where $r_{i,t}$ is the return of stock
$i$ on day $t$, $r_{m,t}$ is the market return, and $\beta_{i,t}$ is
estimated from the preceding 120 trading days. Market volatility
captures the overall level of systematic uncertainty, whereas
idiosyncratic volatility measures stock-specific uncertainty after
removing broad market movements. Together, these variables provide
complementary information about the prevailing market regime. Across
experiments, both state variables are fixed, and only their point of
integration into the model is varied.




\subsection{Task}\label{sec:task}

The task is to predict cross-sectional stock volatility. For stock $i$
on day $t$, the target is the annualized five-day forward realized volatility
\begin{equation}
y_{i,t}
=
\sqrt{252}\,
\operatorname{Std}
\left(
r_{i,t+1},\ldots,r_{i,t+5}
\right),
\label{eq:target}
\end{equation}
where $r_{i,t+k}$ denotes the daily log return of stock $i$ on day
$t+k$. Models generate predictions for all stocks on each trading day.

Evaluation follows the rolling walk-forward protocol illustrated in
Table~\ref{tab:walkforward}. Each evaluation window uses a 504-trading-day (two years)
development period, with the first 85\% used for model training and the
final 15\% used for validation, followed by a 63-trading-day (one quarter) test period.
After each evaluation period, the full window advances by 63 trading
days. This procedure produces 30 non-overlapping test windows from
April 2018 to October 2025 and approximately 1.9 million out-of-sample
forecasts. The evaluation period spans diverse market regimes, including the COVID-19 market crash in 2020, the subsequent recovery in 2021, the broad bear market of 2022, and comparatively calmer periods before and afterward.


\begin{table}[t]
\centering
\caption{Rolling walk-forward evaluation protocol.}
\label{tab:walkforward}
\small
\setlength{\tabcolsep}{10pt}
\begin{tabular}{lcc}
\toprule
Window & Development period & Test period \\
\midrule
1  & 2016-04 -- 2018-03 & 2018-04 -- 2018-06 \\
2  & 2016-07 -- 2018-06 & 2018-07 -- 2018-09 \\
$\vdots$ & $\vdots$ & $\vdots$ \\
30 & 2023-08 -- 2025-07 & 2025-08 -- 2025-10 \\
\bottomrule
\end{tabular}
\end{table}
\section{Model Architecture}\label{sec:models}

All neural network models are built from the same two-hidden-layer MLP
block, which serves as the common forecasting component across all
architectures. The variants differ in three aspects: the use of regime state variables, the routing mechanism, and whether experts predict full volatility forecasts or residual corrections to a shared base forecast.

\subsection{Shared MLP block}\label{sec:cell}

To keep the forecasting component consistent across models, every
neural network is constructed from the same two-hidden-layer MLP block.
Given an input vector $q$, the block is defined as
\begin{equation}\label{eq:cell}
\begin{aligned}
h_1 &= \phi(W_1 q + b_1),\\
h_2 &= \phi(W_2 h_1 + b_2),\\
\operatorname{Block}(q;\theta) &= W_3 h_2 + b_3.
\end{aligned}
\end{equation}
Here, $\phi$ denotes the GELU activation. Both hidden layers have width
$H$, and dropout with rate 0.1 is applied after each hidden layer during
training. Let $x$ denote the stock-level feature vector and $z$ denote
the vector of regime state variables. The MLP baselines receive
either the stock features alone, $q=x$, or the concatenated stock and
state inputs, $q=(x,z)$. In MoE and RG-ResMoE, each forecasting block
receives only $x$, while $z$ can affect the prediction only through the
gate.

\subsection{Standard MoE}\label{sec:moe-std}

A standard mixture-of-experts (MoE) model combines the predictions of
multiple experts, each implemented as an MLP block, through a learned
gate. Intuitively, different experts can specialize in different market
conditions or stock-level patterns, while the gate determines how much
each expert contributes to the final prediction. Given the stock-level
feature vector $x$, each expert produces a volatility forecast, and the
gate combines these forecasts as
\begin{equation}\label{eq:moe}
\begin{aligned}
e_k(x) &= \operatorname{Block}(x;\theta_k),\\
\pi &= \operatorname{softmax}\bigl(g(u)\bigr),\\
\hat y &= \sum_{k=1}^{K} \pi_k e_k(x).
\end{aligned}
\end{equation}
Here, $e_k(x)$ is the volatility forecast produced by the $k$th MLP
expert, $\theta_k$ denotes its parameters, $g$ is the gate network, and
$\pi_k$ is the weight assigned to expert $k$. Each expert receives $x$,
while the gate receives both the stock-level features and the volatility
state variables, so $u=(x,z)$. The gate therefore provides
input-dependent conditional weighting. All experts and the gate are
trained jointly.

\subsection{Regime-gated Residual MoE (RG-ResMoE)}\label{sec:resmoe}

Unlike the standard MoE, in which each expert predicts the full volatility level, RG-ResMoE decomposes the prediction into a base network and expert-specific residual corrections, as illustrated in Figure~\ref{fig:framework}. The base network captures
common forecasting patterns, while the experts model conditional
deviations from the base prediction. This decomposition allows the
experts to focus on specialized corrections rather than relearning the
full forecasting function.

Given the stock-level feature vector $x$, the base network produces
\begin{equation}
\hat{y}_{\mathrm{base}} =\operatorname{Block}(x;\theta_b),
\end{equation}
where $\hat{y}_{\mathrm{base}}$ is the base forecast. The $k$th expert produces a residual
correction,
\begin{equation}
r_k(x)=\operatorname{Block}(x;\theta_k).
\end{equation}
The gate then combines the expert corrections as
\begin{equation}\label{eq:RG-ResMoE}
\begin{aligned}
\pi &= \operatorname{softmax}\bigl(g(u)\bigr),\\
\hat{y}_{\mathrm{correction}}
&= \sum_{k=1}^{K}\pi_k r_k(x),\\
\hat{y}
&= \hat{y}_{\mathrm{base}}+\hat{y}_{\mathrm{correction}}.
\end{aligned}
\end{equation}

Here, $\theta_b$ and $\theta_k$ denote the parameters of the base
network and the $k$th expert, respectively, and $\pi_k$ is the weight
assigned to expert $k$. The aggregated residual correction
$\hat{y}_{\mathrm{correction}}$ is the soft-weighted sum of the expert outputs.
The base network and all experts receive $x$, while the gate receives
$u=(x,z)$.

Training proceeds in two stages. The base network is trained first and
then frozen. The $K=4$ correction experts and the gate are then trained
jointly by minimizing
\begin{equation}\label{eq:loss}
\mathcal{L}
= \operatorname{MSE}(\hat y,y)
+ \alpha\,
\overline{\left(\sum_{k=1}^{K}\pi_k r_k(x)\right)^2}
+ \lambda_{\mathrm{LB}}
\sum_{k=1}^{K}
\left(\bar{\pi}_k-\frac{1}{K}\right)^2.
\end{equation}
Here, the bar denotes an average over the full training batch. The first term is
the prediction loss. The second term penalizes the magnitude of the
aggregate residual correction, encouraging the final prediction to
remain close to the frozen base forecast $\hat{y}_{\text{base}}(x)$. The coefficient
$\alpha$ controls the strength of this shrinkage. The third term is a
load-balancing penalty that discourages the routing weights from
concentrating on a small number of experts. The coefficient
$\lambda_{\mathrm{LB}}$ controls how strongly the routing weights are encouraged toward the uniform allocation $1/K$.
Together, these regularizers keep the expert corrections complementary
to the base forecast and reduce the risk of routing collapse. Each
correction expert uses a zero-initialized final layer, so
$r_k(x)=0$ at initialization and the model initially reproduces the
base forecast $\hat{y}_{\text{base}}(x)$.

\begin{table}[t]
\caption{Models and architectural variants considered in the experiments. +$z$ denotes direct input integration of regime variables, while -$z$ removes the regime pathway from RG-ResMoE.}
\label{tab:models}
\centering
\small
\setlength{\tabcolsep}{5pt}
\begin{tabular}{p{0.29\linewidth}p{0.59\linewidth}}
\toprule
Category & Models \\
\midrule
Main models &
RG-ResMoE \\

Regime state variants &
RG-ResMoE (-$z$); Ridge (+$z$); MLP-S (+$z$); MLP-L (+$z$) \\

Soft routing &
Learned soft gate \\

Hard routing &
Learned top-1 gate; volatility-quantile assignment;
market$\times$idiosyncratic volatility split; GICS sectors \\

Architecture variants &
Standard MoE; random initialization; RG-ResMoE-K2; RG-ResMoE-K6 \\

Baselines &
Persistence; GARCH; HAR; Ridge; MLP-S; MLP-L \\
\bottomrule
\end{tabular}
\end{table}

\subsection{Routing strategies}\label{sec:routing-strategies}

The routing mechanism determines how expert predictions are selected and
combined. We consider both soft and hard routing.
In soft routing, a gate network produces a probability distribution over
the experts, $\pi=\operatorname{softmax}(g(u))$,
and the final prediction is formed as a weighted combination of all
expert outputs. This allows multiple experts to contribute to each
forecast, with their weights varying continuously across observations.

In hard routing, each observation is assigned to a single expert. The
selected expert may be determined either by a learned top-1 gate or by
predefined rules based on market characteristics, such as volatility
regimes or sector membership. The prediction is produced solely by the
selected expert, without averaging across experts.
Both routing strategies use the same expert architecture and training
budget. They differ only in how expert specialization is enforced,
allowing us to isolate the effect of routing independently of model
capacity.

\subsection{Comparison set}\label{sec:gates}

Table~\ref{tab:models} summarizes the models evaluated in this study.
The comparison includes classical statistical baselines, MLP
architectures of different capacities, MoE and the proposed RG-ResMoE family.
Beyond the main RG-ResMoE model, the variants isolate three design
dimensions: the use of regime state variables, the routing
mechanism, and the RG-ResMoE architecture itself. Since all neural models
share the same forecasting block, these comparisons isolate the effect
of each design choice while keeping the underlying predictor
comparable.

\begin{table*}[t]
\caption{Main forecasting results on the U.S. equity panel. Persistence and GARCH are estimated per stock. All other models are pooled across stocks. Neural network results report mean $\pm$ standard deviation over 30 seeds. Shading indicates statistically significant paired differences from RG-ResMoE:
\colorbox{sigA}{\strut $p{<}0.001$},
\colorbox{sigB}{\strut $p{<}0.01$}, and
\colorbox{sigC}{\strut $p{<}0.05$}.}
\label{tab:headline}
\centering
\small
\setlength{\tabcolsep}{7pt}

\begin{tabular}{llcccccc}
\toprule
Training
& Model
& \multicolumn{5}{c}{Forecast performance}
& Stability \\
\cmidrule(lr){3-7}
\cmidrule(lr){8-8}
&
& IC $\uparrow$
& RMSE $\downarrow$
& $R^2$ $\uparrow$
& ICIR $\uparrow$
& QLIKE $\downarrow$
& Collapse $\downarrow$ \\
\midrule

\multirow{2}{*}{Per-stock}
& Persistence
& \cellcolor{sigA}0.5018
& \cellcolor{sigA}0.2570
& \cellcolor{sigA}0.121
& 5.18
& \cellcolor{sigC}0.799
& --- \\

& GARCH(1,1)
& \cellcolor{sigA}0.5200
& \cellcolor{sigA}0.2642
& \cellcolor{sigA}0.070
& 5.48
& \cellcolor{sigC}1.120
& --- \\

\midrule

\multirow{6}{*}{Pooled}
& HAR
& \cellcolor{sigA}0.5315
& \cellcolor{sigA}0.2347
& \cellcolor{sigA}0.266
& 5.93
& 0.749
& --- \\

& Ridge
& \cellcolor{sigA}0.5266
& \cellcolor{sigA}0.2336
& \cellcolor{sigA}0.273
& 5.75
& 0.760
& --- \\

& MLP-S
& \cellcolor{sigA}0.5393\,$\pm$\,0.0034
& \cellcolor{sigA}0.2323\,$\pm$\,0.0011
& \cellcolor{sigA}0.280\,$\pm$\,0.007
& 6.00\,$\pm$\,0.15
& 74.3\,$\pm$\,395.1
& 2/30 \\

& MLP-L
& \cellcolor{sigA}0.5421\,$\pm$\,0.0033
& \cellcolor{sigA}0.2320\,$\pm$\,0.0009
& \cellcolor{sigA}0.282\,$\pm$\,0.005
& 6.08\,$\pm$\,0.18
& 2.23\,$\pm$\,5.15
& 3/30 \\

& Standard MoE
& \cellcolor{sigA}0.5413\,$\pm$\,0.0027
& \cellcolor{sigA}0.2357\,$\pm$\,0.0049
& \cellcolor{sigA}0.259\,$\pm$\,0.031
& 6.07\,$\pm$\,0.14
& \cellcolor{sigC}8{,}775\,$\pm$\,18{,}277
& 24/30 \\

& \textbf{RG-ResMoE (ours)}
& \textbf{0.5469}\,$\pm$\,0.0012
& \textbf{0.2304}\,$\pm$\,0.0013
& \textbf{0.292}\,$\pm$\,0.008
& \textbf{6.14}\,$\pm$\,0.06
& \textbf{0.735}\,$\pm$\,0.017
& \textbf{0/30} \\

\bottomrule
\end{tabular}
\end{table*}

\section{Evaluation Protocol}\label{sec:setup}

\subsection{Training and hyperparameter tuning}\label{sec:training}

All neural network models use the same training procedure: full-batch
Adam optimization of mean-squared error (MSE), early stopping based on
validation loss, and restoration of the best checkpoint. Each reported
configuration is trained with 30 random seeds, while hyperparameter
sweeps use three seeds. This design allows us to evaluate both average
performance and sensitivity to random initialization under a fixed
training budget.

Hyperparameters are selected using validation information coefficient
(IC), defined as the cross-sectional correlation between predicted and
realized volatility. Candidate configurations are filtered using three
robustness criteria. The selected configuration (i) lies in the
interior of the search grid rather than at its boundary, (ii) achieves
a mean validation IC exceeding that of the runner-up by more than its
across-seed standard deviation, and (iii) has a validation MSE within
one standard deviation of the lowest MSE observed in the sweep.

\subsection{Baselines}\label{sec:baselines}

The comparison set includes three classical forecasting baselines:
20-day persistence, pooled HAR \cite{corsi2009simple}, and per-stock
GARCH(1,1) \cite{bollerslev1986generalized}. Ridge provides a linear
baseline using the same stock-level features as the neural networks.
MLP-S (\textbf{small}) uses a single MLP block with hidden width
$H{=}16$, matching an individual block in RG-ResMoE. MLP-L
(\textbf{large}) uses the same block structure with a larger hidden
width $H{=}44$ to match the overall forecasting capacity of the full
RG-ResMoE model.
To distinguish the effect of using the regime variables as predictor
inputs from their use in expert routing, we additionally evaluate
Ridge($+z$), MLP-S($+z$), and MLP-L($+z$). These variants append the two
regime state variables to the ordinary model inputs and are re-tuned
from scratch.

\subsection{Evaluation metrics}\label{sec:metrics}

Our primary accuracy metrics are IC, RMSE, and out-of-sample $R^2$.
We additionally report ICIR, QLIKE, VaR calibration, and the frequency
of collapsed seeds to assess ranking consistency, volatility
calibration, and training stability.

\textbf{Ranking performance:} Daily cross-sectional Spearman
information coefficient (IC) between predicted and realized volatility,
averaged over test days, and information coefficient information ratio (ICIR), defined as the mean daily IC divided
by its time-series standard deviation.

\textbf{Level accuracy:} Root mean-squared error (RMSE) and
out-of-sample $R^2$, pooled across all stock-days and computed on the
annualized volatility scale.

\textbf{Quasi-likelihood loss (QLIKE):} The volatility loss
\begin{equation}
\mathrm{QLIKE}_{i,t}
=
\frac{\sigma_{i,t}^{2}}{\hat{\sigma}_{i,t}^{2}}
-
\ln\left(
\frac{\sigma_{i,t}^{2}}{\hat{\sigma}_{i,t}^{2}}
\right)
-1,
\end{equation}
averaged across all stock-days \cite{patton2011volatility}. Lower values
indicate better variance forecasts.

\textbf{Value at Risk (VaR) coverage:} The fraction of tickers for which the
null of correct unconditional coverage is rejected by Kupiec's test
\cite{kupiec1995techniques} at the 5\% significance level. We evaluate
both 5\% and 1\% VaR, using Student-$t$ residual distributions estimated
from the corresponding training window.

\textbf{Training stability:} The number of collapsed seeds,
defined using a prespecified diagnostic threshold as runs whose mean
test QLIKE exceeds 2.0.


\subsection{Statistical testing}\label{sec:stats}

Pairwise Diebold--Mariano-style tests
\cite{diebold1995comparing} assess whether the forecast performance of
two competing models differs significantly. For each test day, we
compute the difference in the selected metric between models $A$ and
$B$ using predictions on the same stocks, then average this difference
across the $S$ random seeds:
\begin{equation}\label{eq:dm}
d_t =
\frac{1}{S}\sum_{s=1}^{S}
\left[X_{A,s}(t)-X_{B,s}(t)\right].
\end{equation}
The resulting series of approximately 1{,}900 daily differences is
tested using Newey--West standard errors
\cite{newey1987simple} with four lags to account for dependence induced
by the overlapping five-day targets. We report the resulting $p$-values alongside the model performance
metrics.



\section{Main Results}\label{sec:results}

Table~\ref{tab:headline} summarizes the main forecasting results.
Across all pooled models, RG-ResMoE achieves the best overall forecast
performance. It obtains the highest IC, lowest RMSE,
highest out-of-sample $R^2$, highest ICIR, and lowest
QLIKE. Most differences relative to the competing neural networks
are statistically significant.

Compared with the classical baselines, all pooled neural network models improve
forecast accuracy substantially, demonstrating the benefit of learning
shared representations across stocks. Within the pooled models, however,
simply increasing model capacity provides only modest gains. Although
MLP-L has approximately the same overall capacity as RG-ResMoE, it remains
consistently worse across every forecast metric. This indicates that the
improvement comes primarily from the residual expert architecture rather
than from additional parameters.

The optimization results further support this conclusion. Standard MoE
shows frequent optimization failures, collapsing in 24 of the 30 random
seeds, whereas RG-ResMoE completes all runs successfully without a single
collapse. The \textbf{residual decomposition} improves not only
forecast accuracy but also training stability.

\section{Where the Gain Comes From}\label{sec:mechanism}

\begin{table}[t]
\caption{Effect of regime-information integration pathway.}
\label{tab:pathway}
\centering
\small
\setlength{\tabcolsep}{6pt}
\begin{tabular}{@{}llccc@{}}
\toprule
Model & Pathway & IC $\uparrow$ & $\Delta$IC $\uparrow$ & Collapse $\downarrow$\\
\midrule
Ridge & --- & 0.5266 & ref. & --- \\
Ridge (+$z$) & input & 0.5243 & $-0.0023$ & --- \\
\midrule
MLP-S & --- & 0.5393 $\pm$ 0.0034 & ref. & 2/30 \\
MLP-S (+$z$) & input & 0.5383 $\pm$ 0.0041 & $-0.0011$ & 22/30 \\
\midrule
MLP-L & --- & 0.5421 $\pm$ 0.0033 & ref. & 3/30 \\
MLP-L (+$z$) & input & 0.5378 $\pm$ 0.0035 & $-0.0043$ & 30/30 \\
\midrule
RG-ResMoE (-$z$) & --- & 0.5466 $\pm$ 0.0011 & ref. & 1/30 \\
RG-ResMoE & gate & \textbf{0.5469} $\pm$ 0.0012 & \textbf{+0.0004} & \textbf{0/30} \\
\bottomrule
\end{tabular}
\end{table}

\subsection{Where should regime information enter the model?}
\label{sec:pathway}

RG-ResMoE uses the two regime state variables only for expert routing,
rather than treating them as ordinary forecasting inputs. This design
raises a natural question: does the improvement come from the regime
information itself, or from how that information is incorporated into
the model?

\textbf{Input integration versus gate-based integration.}
Table~\ref{tab:pathway} compares two ways of using the same state
variables. In the input pathway, the state variables are appended to
the stock-level features used by Ridge, MLP-S, and MLP-L, allowing
them to directly influence the forecast. In the gate pathway, the
forecasting networks continue to use only the stock-level features,
while the state variables are used exclusively by the RG-ResMoE gate to
determine the mixture of residual experts.

The two integration strategies lead to markedly different outcomes.
Appending the state variables to the forecasting inputs consistently
reduces IC and substantially increases training instability for the
MLP baselines. In contrast, using the same variables only for expert
routing produces a small improvement in IC while eliminating the
remaining collapsed run. These results indicate that the value of
regime information lies not in expanding the predictor input, but in
adapting how residual corrections are combined. Gate-based integration
allows the state variables to modulate the expert weights without
directly shifting the forecast produced by the underlying forecasting
networks.

\begin{figure}[t]
\centering
\includegraphics[width=0.85\linewidth]{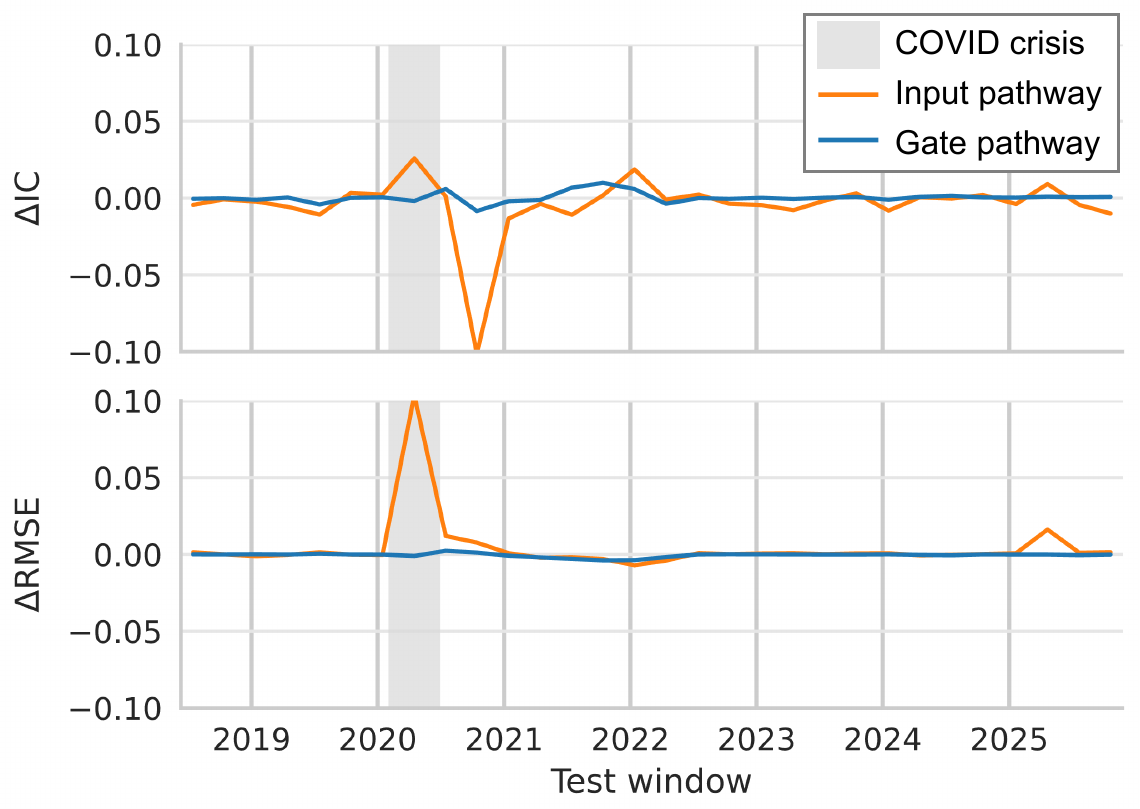}
\caption{Per-window comparison of input and gate integration across the 30 walk-forward windows. Lines show seed-averaged paired differences within each evaluation window. Positive values indicate improvement for $\Delta$IC and deterioration for $\Delta$RMSE.}
\label{fig:drift}
\end{figure}

\textbf{Behavior across market regimes.}
The aggregate results in Table~\ref{tab:pathway} may conceal variation
across market conditions. Figure~\ref{fig:drift} therefore reports
window-level differences between models with and without regime state
variables. For the input pathway, the difference is computed as
MLP-L(+$z$) minus MLP-L; for the gate pathway, it is computed as
RG-ResMoE minus RG-ResMoE(-$z$). Positive values favor the model with
regime variables for IC, whereas negative values favor it for RMSE. Positive values indicate improvement for
$\Delta$IC and deterioration for $\Delta$RMSE.

The largest difference between the two pathways appears around the
transition from the COVID shock to the subsequent market recovery.
During the stress period, the input pathway briefly improves IC, but
this gain reverses during the recovery phase. RMSE also increases
substantially during the stress period before moving back toward zero.

The gate pathway exhibits much smaller changes over the same period
and does not show the pronounced IC reversal observed under direct
input integration. This comparison suggests that the effect of using
regime variables as forecasting inputs is highly dependent on market
conditions. Using the same information through the gate instead
produces a more stable response during major regime transitions.

\begin{figure}[t]
\centering
\includegraphics[width=0.8\linewidth]{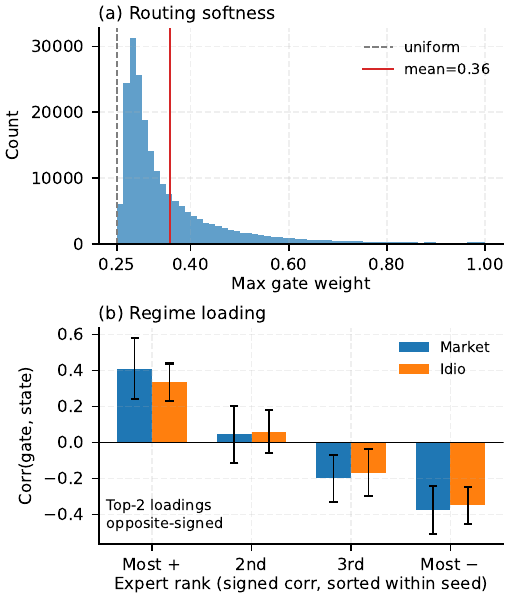}
\caption{
Gate behavior of RG-ResMoE across 30 random seeds. (a) Distribution of the maximum routing weight assigned to any expert for each observation. Uniform routing corresponds to 0.25 and hard routing to 1. (b) Correlations between expert weights and the two regime variables after within-seed sorting. Whiskers denote across-seed standard deviations.}
\label{fig:softness}
\end{figure}

\subsection{What does the gate learn?}
\label{sec:gate-learning}

The learned routing remains broadly distributed across experts rather than collapsing to a single component (Figure~\ref{fig:softness}a). The average maximum gate weight remains close to uniform routing, and only a small fraction of observations are assigned almost entirely to one expert. At the same time, expert weights show substantial sensitivity to both market and idiosyncratic volatility, with the most strongly loaded experts typically exhibiting opposite-signed correlations (Figure~\ref{fig:softness}b). Together, these patterns indicate that the gate performs continuous state-dependent reweighting rather than imposing discrete regime assignments.

\begin{table}[t]
\caption{Comparison of soft and hard routing. All variants use the same frozen base and four residual experts. $\Delta$IC and $p$-values are reported relative to RG-ResMoE.}
\label{tab:routing}
\centering
\small
\setlength{\tabcolsep}{2pt}
\begin{tabular}{@{}llcccc@{}}
\toprule
Variant & Routing & IC $\uparrow$ & $\Delta$IC $\uparrow$ & $p$ & Collapse $\downarrow$\\
\midrule
RG-ResMoE & soft & \textbf{0.5469 $\pm$ 0.0012} & ref. & & \textbf{0/30} \\
learned top-1 & hard & 0.5448 $\pm$ 0.0018 & $-0.0021$ & $<10^{-4}$ & 1/30 \\
volatility quantiles & hard & 0.5447 $\pm$ 0.0015 & $-0.0022$ & $<10^{-4}$ & 1/30 \\
GICS sectors & hard & 0.5445 $\pm$ 0.0016 & $-0.0024$ & $<10^{-4}$ & 1/30 \\
market $\times$ idio split & hard & 0.5435 $\pm$ 0.0016 & $-0.0034$ & $<10^{-4}$ & 1/30 \\
\bottomrule
\end{tabular}
\end{table}

\subsection{Is soft routing better than hard routing?}
\label{sec:routing}

The previous analysis shows that the gate learns a continuous
redistribution of expert weights. An immediate question is whether
this continuous routing is necessary, or whether comparable
performance can be achieved using discrete expert assignments.

Table~\ref{tab:routing} compares the learned soft gate with four
hard-routing alternatives while keeping the residual architecture and
the number of experts fixed. The hard variants include a learned
top-1 gate, volatility quantile assignment, GICS sectors,
and a market-volatility $\times$ idiosyncratic-volatility split.
All hard-routing variants underperform the soft gate. Their IC ranges
from 0.5435 to 0.5448, corresponding to reductions of
$0.0021$--$0.0034$ relative to RG-ResMoE, with every comparison
significant at $p<10^{-4}$. The performance degradation is consistent
across all four routing strategies rather than being attributable to a
particular hand-crafted partition.
These results indicate that the improvement does not arise from a
better definition of market regimes. Instead, allowing multiple
experts to contribute simultaneously is consistently more effective
than assigning each observation to a single expert.

\begin{table}[t]
\caption{Ablations on expert count and residual design. $\Delta$IC is relative to the main RG-ResMoE model.}
\label{tab:design}
\centering
\small
\setlength{\tabcolsep}{5.5pt}
\begin{tabular}{lccc}
\toprule
Variant & IC $\uparrow$ & $\Delta$IC $\uparrow$ & Collapse $\downarrow$ \\
\midrule
RG-ResMoE-k4 (main) &
\textbf{0.5469 $\pm$ 0.0012} &
ref. &
\textbf{0/30} \\

RG-ResMoE-K2 &
0.5466 $\pm$ 0.0014 &
$-0.0004$ &
0/30 \\

RG-ResMoE-K6 &
0.5467 $\pm$ 0.0012 &
$-0.0002$ &
3/30 \\

RG-ResMoE (random init) &
0.5430 $\pm$ 0.0021 &
$-0.0039$ &
13/30 \\

Standard MoE &
0.5413 $\pm$ 0.0027 &
$-0.0056$ &
24/30 \\
\bottomrule
\end{tabular}
\end{table}


\subsection{How many experts are needed?}
\label{sec:minimal}

Two aspects of the expert architecture are examined in
Table~\ref{tab:design}: the number of experts and the residual
parameterization. To isolate these effects, the expert-count variants
maintain approximately the same overall model capacity by widening
individual experts when fewer are used and narrowing them when more
experts are introduced.

The number of experts has only a modest effect on overall forecasting
accuracy. Reducing the architecture from four experts to two lowers IC
by only 0.0004, indicating that most of the benefit is retained with a
smaller mixture. Increasing to six experts provides no further
improvement, while substantially reducing training stability. Although
the average IC remains similar to the four-expert model, three of
thirty runs collapse and QLIKE becomes considerably more variable.

The residual parameterization is substantially more important.
Replacing the zero-initialized residual layer with random
initialization lowers IC and increases the number of collapsed runs
from 0 to 13. Removing the frozen base forecast entirely has an even
larger effect. The resulting standard MoE exhibits the lowest IC among
the neural mixture models and collapses in 24 of 30 runs.

These results indicate that the performance gain does not arise from a
large ensemble of experts. Instead, the principal contribution comes
from the residual architecture, which anchors expert learning around a
shared base forecast while allowing gated residual corrections to model
state-dependent deviations.

\section{Practical Evaluation}\label{sec:practical}

Average forecast accuracy does not fully characterize the practical
value of a forecasting model. We therefore evaluate RG-ResMoE from three
deployment-oriented perspectives: training reliability, risk
calibration, and performance during periods of market stress.

\begin{figure}[t]
\centering
\includegraphics[width=0.85\linewidth]{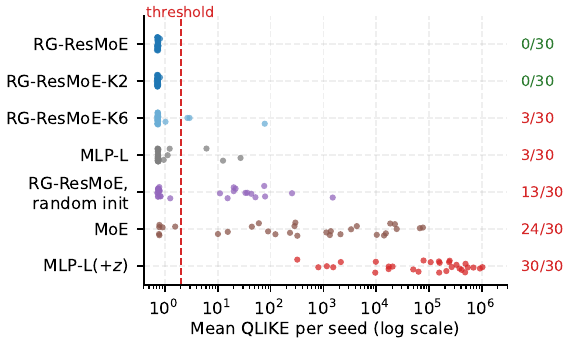}
\caption{Training stability across random initializations. Each point shows the mean test QLIKE for one training seed. The dashed line marks the collapse threshold (QLIKE = 2), and the numbers on the right report the number of collapsed runs.}
\label{fig:collapse}
\end{figure}

\textbf{Training reliability.}
Figure~\ref{fig:collapse} summarizes training reliability. RG-ResMoE and RG-ResMoE-K2 complete all 30 runs without collapse, while RG-ResMoE-K6 and MLP-L each collapse in 3 of 30 runs. Random initialization and standard MoE are substantially less stable, with 13 and 24 collapsed runs, respectively, confirming that the residual design and frozen base are the primary sources of robustness.


\begin{table}[t]
\caption{Kupiec VaR coverage rejection rates averaged over 30 seeds. Lower values indicate better calibration. $p$-values are from paired tests with RG-ResMoE as the reference. n.s. denotes not significant.}
\label{tab:var}
\centering
\small
\begin{tabular}{lccc}
\toprule
Slice & RG-ResMoE & MLP-L & MLP-S \\
\midrule
5\% VaR, all dates & \textbf{26.7\%} & 30.3\% ($p{<}10^{-4}$) & 32.7\% ($p{=}.0007$) \\
5\% VaR, high-vol  & \textbf{11.9\%} & 14.8\% ($p{<}10^{-4}$) & 16.9\% ($p{=}.0001$) \\
5\% VaR, low-vol   & \textbf{22.9\%} & 24.7\% ($p{=}.003$)    & 28.2\% ($p{=}.002$) \\
1\% VaR, all dates & \textbf{9.1\%}  & 9.9\% ($p{=}.001$)     & 11.4\% ($p{=}.0001$) \\
1\% VaR, high-vol  & \textbf{6.2\%}  & 7.1\% ($p{<}10^{-4}$)  & 7.2\% ($p{<}10^{-5}$) \\
1\% VaR, low-vol   & 5.8\%           & 5.8\% (n.s.)           & 6.8\% ($p{=}.053$, n.s.) \\
\bottomrule
\end{tabular}
\end{table}

\textbf{Risk calibration.} Besides producing accurate forecasts, a
volatility model should also provide well-calibrated estimates of
downside risk. We therefore convert each five-day volatility forecast
into one-day Value-at-Risk (VaR) thresholds, which estimate the loss
level expected to be exceeded only with a specified probability (e.g.,
5\% or 1\%).

Table~\ref{tab:var} reports the fraction of tickers rejected by
Kupiec's coverage test, where lower values indicate better calibrated
VaR estimates. RG-ResMoE consistently achieves the lowest rejection rates
across nearly all settings, significantly outperforming both MLP
baselines in five of the six regime--VaR combinations. The largest
improvements occur during high-volatility periods, when accurate
tail-risk estimation is most critical for risk management. The only
exception is the 1\% low-volatility slice, where all models perform
similarly and the differences are not statistically significant.

\begin{table}[t]
\caption{RG-ResMoE’s IC advantage over MLP-L across market conditions. Relative gain is the ratio of each subset’s $\Delta$IC to the full-sample $\Delta$IC.}
\label{tab:stress}
\centering
\small
\begin{tabular}{lcccc}
\toprule
Slice & Days & $\Delta$IC $\uparrow$ & Rel. gain $\uparrow$ & $p$ \\
\midrule
Full sample                     & 1{,}890 & $+0.0048$ & ref. & $<10^{-4}$ \\
Top market-vol decile           & 189     & $+0.0207$ & $4.3\times$ & $<10^{-4}$ \\
COVID crisis (2020-02--06)      & 104     & $+0.0322$ & $6.7\times$ & $<10^{-4}$ \\
2022 bear market                & 251     & $+0.0079$ & $1.6\times$ & $<10^{-4}$ \\
Regime-flip windows ($\pm$10d)  & 826     & $+0.0031$ & $0.6\times$ & $<10^{-4}$ \\
\bottomrule
\end{tabular}
\end{table}

\textbf{Performance under market stress.}
We finally examine where RG-ResMoE's ranking advantage is concentrated,
particularly during periods of elevated market volatility. For each
market condition, Table~\ref{tab:stress} computes RG-ResMoE's IC
advantage over MLP-L using only the dates in that subset.

The gain is not uniform over time. RG-ResMoE outperforms MLP-L by
$+0.0048$ IC in the full sample, but by $+0.0207$ IC in the
highest-volatility decile and $+0.0322$ IC during the COVID crisis,
corresponding to approximately $4.3\times$ and $6.7\times$ the
full-sample gain. The 2022 bear market also amplifies the advantage,
although less strongly. By contrast, regime-transition windows do not
show a similar amplification. These results suggest that the benefit
of residual routing is greatest during sustained periods of elevated
market volatility rather than during regime transitions themselves.


\begin{table}[t]
\caption{Replication on the Japanese TSE Prime market.}
\label{tab:japan}
\centering
\small
\setlength{\tabcolsep}{2pt}
\begin{tabular}{@{}lcccc@{}}
\toprule
Model & IC $\uparrow$ & RMSE $\downarrow$ & QLIKE $\downarrow$ & Collapse $\downarrow$ \\
\midrule
MLP-L & \cellcolor{sigA}0.4815\,$\pm$\,0.0036 & \cellcolor{sigA}0.1851\,$\pm$\,0.0007 & \cellcolor{sigB}11.6\,$\pm$\,34.2 & 6/30 \\
MLP-L (+$z$) & \cellcolor{sigA}0.4815\,$\pm$\,0.0026 & \cellcolor{sigB}0.1901\,$\pm$\,0.0025 & 41{,}256\,$\pm$\,40{,}296 & 28/30 \\
MoE & \cellcolor{sigA}0.4805\,$\pm$\,0.0029 & \cellcolor{sigA}0.1862\,$\pm$\,0.0012 & 3{,}081\,$\pm$\,7{,}227 & 22/30 \\
\textbf{RG-ResMoE} & \textbf{0.4858\,$\pm$\,0.0024} & \textbf{0.1846\,$\pm$\,0.0004} & \textbf{0.80\,$\pm$\,0.33} & \textbf{1/30} \\
\bottomrule
\end{tabular}
\end{table}

\section{Cross-Market Replication}\label{sec:japan}

We evaluate generalization beyond the U.S. market using the Japanese
TSE Prime panel. Table~\ref{tab:japan} reports the core comparison
under the same features, walk-forward protocol, hyperparameter
selection rules, and thirty random seeds. RG-ResMoE outperforms MLP-L by
$+0.0043$ IC (IC and RMSE $p<10^{-4}$; QLIKE $p=0.002$). Input
concatenation is again unstable. It leaves IC unchanged while
worsening RMSE and collapsing in 28 of 30 MLP-L (+$z$) seeds. The
same stability ordering is reproduced on the Japanese market,
suggesting that the benefit of residual routing is not
market-specific.

\section{Related Work}\label{sec:related}

\textbf{Regime switching in finance.} Markov-switching models made
regime dependence central to macroeconomic and volatility modeling
\cite{hamilton1989new,hamilton1994autoregressive,gray1996modeling}.
Recent neural forecasting methods extend this idea through
regime-aware architectures and expert routing
\cite{bildirici2014modeling,yu2024miga,he2026raven}.
These approaches demonstrate the value of incorporating market
regimes, but typically modify the regime representation, routing
strategy, and model capacity simultaneously. As a result, the contribution of the integration point remains unclear.

\textbf{Mixture-of-experts.} MoE models originated with \citet{jacobs1991adaptive} and have
become a standard architecture for scalable conditional computation
\cite{shazeer2017outrageously,fedus2022switch}, including recent
time-series foundation models
\cite{shi2025timemoe,liu2025moirai}. Modern sparse MoE architectures
primarily use hard top-$k$ routing to increase model capacity while
keeping the computation per input nearly constant, with expert
specialization emerging through conditional activation. Soft expert
combinations have also been studied to improve optimization and
training stability \cite{puigcerver2024soft}. However, most of this literature treats routing as a mechanism for computational scaling rather than as a modeling choice for incorporating market state information.

\textbf{Residual learning and stable initialization.} Residual
networks improve optimization by learning corrections rather than
complete transformations \cite{he2016deep}. Zero- or near-zero-initialized residual branches further stabilize training
\cite{zhang2019fixup}, while warm-starting methods show that
remaining close to a stable initialization can improve generalization
\cite{ash2020warmstarting}. Although these principles are well
established in deep learning, their combination with gated expert routing has received comparatively little attention, particularly in financial forecasting.

\section{Conclusion and Limitations}\label{sec:conclusion}

This paper studied how regime information should be incorporated into
a neural network for volatility forecasting. The answer is not simply to
add regime variables or adopt an MoE architecture. In our setting, the
same state variables reduce performance when concatenated to the
forecasting network but improve it when used only to gate residual
correction experts. Soft residual correction on top of a frozen base
forecast consistently outperforms hard routing, richer hand-crafted
regime definitions, additional experts, and a standard MoE without a
frozen base.

The broader implication is that nonstationary market information is
most useful for modulating the forecasting process rather than
directly determining the forecast itself. More generally, the results
suggest that the value of MoE in compact forecasting models lies less
in increasing model capacity than in controlling how auxiliary state
information influences the prediction. 

This study is limited to
volatility forecasting in two equity markets using a compact MLP
backbone. Whether the same design principle extends to other financial
forecasting tasks, larger sequence models, or alternative forms of
market state remains an important direction for future work.


\bibliographystyle{ACM-Reference-Format}
\bibliography{references}

\end{document}